\def\year{2022}\relax
\documentclass[letterpaper]{article} 
\usepackage{aaai22}  
\usepackage{times}  
\usepackage{helvet}  
\usepackage{courier}  
\usepackage[hyphens]{url}  
\usepackage{graphicx} 
\usepackage{natbib}  
\usepackage{caption} 
\DeclareCaptionStyle{ruled}{labelfont=normalfont,labelsep=colon,strut=off} 
\usepackage{algorithm}
\usepackage{algorithmic}

\usepackage{newfloat}
\usepackage{listings}
\floatstyle{ruled}
\newfloat{listing}{tb}{lst}{}
\floatname{listing}{Listing}
\usepackage{graphicx}
\usepackage{sgame}
\usepackage{booktabs}
\usepackage{amsfonts}
\usepackage{mathtools}
\usepackage{amsthm}

\DeclareMathOperator*{\argmax}{arg\,max}

\theoremstyle{plain}
\newtheorem{theorem}{Theorem}

\newtheorem{proposition}[theorem]{Proposition}

\theoremstyle{definition}
\newtheorem{definition}[theorem]{Definition}
\newtheorem{remark}[theorem]{Remark}

\newcommand\numberthis{\addtocounter{equation}{1}\tag{\theequation}}

\title{Simulation of Attacker Defender Interaction in a Noisy Security Game}
\author{
    Erick Galinkin,
    Emmanouil Pountourakis,
    John Carter,
    Spiros Mancoridis
}
\affiliations{
    Drexel University\\
    Philadelphia, Pennsylvania 19104\\
    eg657@drexel.edu
}

\begin{document}

\maketitle

\begin{abstract}
In the cybersecurity setting, defenders are often at the mercy of their detection technologies and subject to the information and experiences that individual analysts have.
In order to give defenders an advantage, it is important to understand an attacker's motivation and their likely next best action.
As a first step in modeling this behavior, we introduce a security game framework that simulates interplay between attackers and defenders in a noisy environment, focusing on the factors that drive decision making for attackers and defenders in the variants of the game with full knowledge and observability, knowledge of the parameters but no observability of the state (``partial knowledge''), and zero knowledge or observability (``zero knowledge'').
We demonstrate the importance of making the right assumptions about attackers, given significant differences in outcomes. 
Furthermore, there is a measurable trade-off between false-positives and true-positives in terms of attacker outcomes, suggesting that a more false-positive prone environment may be acceptable under conditions where true-positives are also higher.
\end{abstract}

\section{Introduction}
The use of artificial intelligence in cybersecurity holds tremendous promise, as time is of the essence in dealing with an intrusion.
However, while tasks like malware detection~\cite{ucci2019survey, Raff2017MalwareEXE} and network anomaly detection~\cite{fernandes2019comprehensive, carter2021evaluation} have seen substantial progress, the automation of mitigation and response remains a challenge.
Part of this challenge lies in the fact that there are many possible ways to represent highly heterogeneous security data, while other challenges involve finding appropriate responses to particular events. 
In general, defenders seek to find a response that has a minimal effect on the legitimate use of the information systems they are tasked with defending, while being maximally impactful to potential attackers. 
This sort of min-max problem can be viewed through a game theoretic lens.

In the development of game theoretic models, there is a fundamental tension between constructing a model that is both simple enough to be tractable and descriptive enough to be useful. 
In cybersecurity, this tension is exacerbated due to the high-stakes associated with decisions in conjunction with the fact that the generation, transmission, and response to information happens rapidly and with little human intervention such that on-system activity is effectively instantaneous.
In these games, a common simplifying assumption is that defenders are operating against some adversary that operates according to a fixed strategy~\cite{Liang2013GameSecurity, manshaei2013game}, often represented using the SIRE model from epidemiology~\cite{Khouzani2012Saddle-pointAttack, liu2020novel, tambe2011security}.
In practice, these sorts of adversaries are rare, and there are human decision-making aspects to how an attacker behaves within a victim network.

Understanding attacker decision-making is crucial for the mitigation of attacks.
Ideally, defenders desire near real-time responses to attacker activity as a way to minimize an attacker's ability to achieve their goals. 
In order to generate these responses, it is beneficial to predict the next best action for the attacker and prevent that action from being taken. 
This necessitates modeling both the attacker's uncertainty about the defender -- not knowing what technologies are available to the defender or what actions they are likely to take -- and the defender's uncertainty about the attacker.
To the best of our knowledge, ours is the first work that prioritizes the attacker's decision-making process and places it within a game theoretic framework.

We leverage the framework of Stochastic Bayesian Games~\cite{albrecht2013game} (SBG) as our foundation and consider three different models of attacker knowledge: full knowledge about the state, actions, history, and parameter space; partial knowledge, where the state is hidden but the parameters of the environment are known; and zero-knowledge, where the attacker knows nothing but the actions they have taken and what actions they can take.
In prior work, the full state and action space is always known to all players, as well as the history of play -- only the strategies and utility function, which are determined by the type of the attacker are unknown.
The modification to include hidden information necessitates some adaptation of the most common solution method for Stochastic Bayesian Games: Harsanyi-Bellman ad hoc coordination, which assumes that information about the states and actions is public knowledge.
A key contribution of this work is the development of direct solutions to a partially observable SBG with limited state and action spaces given some conditions on the environment.

In order to study the impact of attacker knowledge, we begin by considering our model with a highly restricted state and action space and simulate attacker interactions with the model.
We find that these scenarios show that a full-knowledge assumption yields much higher expected outcomes for attackers -- an assumption that demands a more aggressive response from defenders.
However, in limited-knowledge scenarios, attackers must conduct some level of reconnaissance or training against a target to estimate their ability to operate, otherwise they perform very poorly.
Moreover, we find that attackers outcomes are largely indifferent to overall alert rates when the rate of detecting malicious activity is high, suggesting that there is a quantifiable trade off against alerts unrelated to malicious activity.

\section{Related Work}
Recent challenges like CAGE~\cite{ttcp2021cage} have encouraged development of models like CybORG~\cite{foley2022autonomous} that use reinforcement learning to train autonomous agents that defend against cyber attacks.
The Ph.D thesis of Campbell~\cite{campbell2022autonomous} also considers a very similar problem space and solution.
These models approach the same problem we wish to consider -- the development of a defensive agent that disrupts an adversary while minimizing impact to network users.
Our work approaches a similar concept from first principles and contributes observations and insights that emerge even in simple models, particularly from the attacker's perspective.

This paper builds primarily on prior work by Albrecht and Ramamoorthy~\cite{albrecht2013game} on Stochastic Bayesian Games (SBG), and the Ph.D dissertation of Maccarone~\cite{maccarone2021stochastic} on applications of SBG to nuclear power plant cybersecurity.
While many security games use Nash equilibria or some comparable equilibrium concept, the complexity of cybersecurity means that often, there are many equilibria and due to hidden information, it is unlikely that all players will identify the same equilibrium.
HBA relies on players selecting actions according to the observed history of the game and some set of decision criteria as the game is played and is not ``predictive'' in the sense of \textit{e.g.}, Nash equilibrium.
We expand the scope of Maccarone's work using the HBA concept in a more general cybersecurity landscape.

The partial observability of the proposed SBG draws on the work of Tom\'{a}\v{s}ek, Bo\v{s}ansk\'{y}, and Nguyen~\cite{tomavsek2020using} on one-sided partially observable stochastic games. 
Their work considered sequential attacks on some target and develops scalable algorithms for solving zero-sum security games in this setting.
These algorithms consider upper and lower value bounds on a subgame, and we couple this partial observability with the aforementioned SBG to more closely approach a real-world setting.

\section{Game Model} \label{sec:game}
\begin{definition} \label{def:sbg}
A Partially Observable Stochastic Bayesian security game is defined as:
\begin{itemize}
    \item $S$, the state space, with initial state $s^0$ and terminal states $\overline{S} \subseteq S$
    \item $N = \{\alpha, \delta\}$, the players of the game: in our case, the attacker $\alpha$ and the defender $\delta$. 
    For each player $i$, we define:
    \begin{itemize}
        \item action space $A_i$
        \item type space $\Theta_i$
        \item utility function $u_i: S \times A \times S \rightarrow \mathbb{R}$
        \item strategy $\pi_i: \mathbb{H} \times A_i \times \Theta_i \rightarrow [0, 1]$
    \end{itemize}
    \item $T: S \times A \times S \rightarrow [0, 1]$, the state transition function
    \item $\Delta: \mathbb{N}_0 \times \Theta \rightarrow [0, 1]$, the distribution of types
    \item $p \coloneqq [0, 1] \times A_{\alpha}$ The alert probability vector 
\end{itemize}
\end{definition}

The element $\mathbb{H}$ contains the history of the game, a sequence of prior states and actions. 
Given the speed at which actions occur, we assume that the attacker and defender move simultaneously at each time step, and let the action $a^t = (a_{\alpha}^t, a_{\delta}^t)$ denote the joint action taken at time $t$.
The history $H^t = \langle s^0, a^0, s^1, a^1, ..., s^t\rangle$ is a concatenation of all states and actions taken from time $0$ until time $t$.
Our type distributions, $\Delta$ are static in the sense that each player has the same type throughout all time steps of the game. 
In our game, the transition function $T$ is always not known by the players, though in some of the discussed cases, parameters of $T$ may be known.

The type space $\Theta$ is the type space of Harsanyi~\cite{harsanyi1967games}, meaning that a player's utility and strategies are determined by their type.
We assume that strategies for particular types are known, despite a player's type being known only to themselves.
In alignment with Albrecht and Ramamoorthy~\cite{albrecht2013game}, a type in this context can be thought of as a ``program'' that governs a player's behavior. 
In the cybersecurity context, we can also contextualize the player's type as the sort of adversary we are playing against. 
This can be defined at a high level \textit{e.g.}, $\Theta_{\alpha} = \{$nation-state, cybercriminal, insider threat$\}$ or at a low level \textit{e.g.}, $\Theta_{\alpha} = \{$APT28, LockBit, Greg, ...$\}$.
A highly granular $\Theta$ will offer more specificity, but increases the complexity of playing the game.

\begin{remark}
Although the model presented in Definition~\ref{def:sbg} contains a single attacker and a single defender, the model can support multiple attackers at the cost of additional state space complexity, since each attacker has their own unique values for the hidden information.
Throughout this paper, all examples assume there is only a single attacker.
\end{remark}

\begin{remark}
We use the term ``know'' to describe information that a player believes with probability 1.
Thus, when considering $H$, each player may have beliefs -- even very strong ones -- about the other player's actions, but knows only the actions they have taken and the elements of $s^t$ that they are able to directly observe.
It is this feature that makes the game ``partially observable'' in contrast to conventional SBG where it is assumed that states and actions are common knowledge.
\end{remark}
\subsection{Harsanyi-Bellman Ad Hoc Coordination}
Ad hoc coordination is derived from the notion of private information in Bayesian games~\cite{harsanyi1967games} coupled with the inclusion of state, probabilistic state transition, and time from Stochastic games~\cite{Shapley1953StochasticGames}.
If every player in a HBA knows the type distribution exactly, then the game admits a Bayesian Nash equilibrium~\cite{harsanyi1968games}.
Ad hoc coordination then, is based on the assertion that each player does not know the type space $\Theta_j$ of the other players and is thus ignorant of $\Delta$.
Since the players are ignorant of $\Delta$, they may identify different posterior distributions and cannot guarantee convergence to a Nash equilibrium~\cite{dekel2004learning}, even a sub-optimal one.
Thus, Albrecht, and Ramamoorthy~\cite{albrecht2013game} use a best-response rule that maximizes expected utility with respect to each player's belief about the type of the other players.

\begin{definition} \label{def:hba}
Harsanyi-Bellman Ad Hoc Coordination (HBA) is defined as 
\[a_i^t \sim \argmax_{a_i} E_{s^t}^{a_i}\left[H^t\right]\] 
where:
\begin{equation} \label{eqn:hba}
\begin{split}
E_{s}^{a_i}\left[\hat{H}\right] & = \sum_{\theta_{-i}^* \in \Theta_{-i}^*} Pr(\theta_{-i}^* | H^t) \\ &\quad \sum_{a_{-i} \in A_{-i}} Q_s^{a_{i, -i}}(\hat{H}) \\ &\quad \prod_{j\neq i}\pi_j(\hat{H}, a_j, \theta_{j}^*)    
\end{split}
\end{equation}
\end{definition}
is the expected long-term payoff for player $i$ taking action $a_i$ in state $s$, given history up to time $t$ $H^t$ and 
\begin{equation}
\begin{split}
    Q_s^a(\hat{H}) & = \sum_{s' \in S} T(s, a, s') \huge( u_i(s, a, s') \\ & + \gamma \max_{a_i} E_{s'}^{a_i} \left[ \langle \hat{H}, a, s' \rangle \right] \huge)
\end{split}
\end{equation}
is the expected long-term payoff for player $i$ when the joint action $a$ is taken in state $s$ and discount factor $\gamma$.
We denote the ``projected history'' of the game, $\hat{H}$, to be the entire history of the game from $0 \leq \tau \leq t$ including the current time step, $t$, assuming that a particular action is taken and concatenated to the history.
This solution concept is very effective in conventional SBG settings, but under our restrictions of partial observability, there are components of $s$ and $a$ that are not seen by both players, creating uncertainty about $H$ for players at any time step. 
We propose non-HBA solutions for our simplified settings in their respective sections.

\section{Simplified Setting} \label{sec:setting}
In our simplified game, we assume an attacking agent and a defending agent who choose their moves simultaneously.
The game is played on an infinite graph, and the attacker's objective is to control as many nodes as possible.
In this game our state space $S$ is a tuple of two integers: one that serves as a count of ``alerts'' in the environment and a second that consists of the number of nodes the attacker has infected.
The action space for defenders consists of two actions: a pass action, where they do nothing, and a shutdown action, that ends the game.
The attacker's action space consists of two actions: infecting an adjacent node and ending the game.

The game is characterized by:
\begin{itemize}
	\item $p$, the probability that an alert occurs given that the attacking player has taken an action.
	\item $q$, the probability that an alert occurs whether or not the attacking player has taken an action.
	\item $Th$, the defender-selected threshold of alerts that ends the game, selected before play begins.
\end{itemize}
Let $p$ and $q$ parameterize two Bernoulli random variables, $X_m \sim \text{Bernoulli}(p)$, $X_n \sim \text{Bernoulli}(q)$.
We define the indicator function 
\[ 
\mathbf{1}_{X} = 
\begin{cases}
	1~&{\text{ if }} X_m = 1 \text{ or } X_n = 1\\
	0~&{\text{ if }} X_n = 0 \text{ and } X_n = 0
\end{cases} 
\]
Let $s_A$ be the total number of alerts that have occurred.
Formally, $s_A = \sum_{i=1}^{t} \mathbf{1}_{X}$. 
This random variable parameterized by $p$ and $q$ also represents part of our transition function $T$, in that the resultant state $s' = (s_A, s_I)$ follows from the change in $s_A$.

If the attacking player ends the game before the defending player, they get a reward, equal to $s_I$, the number of nodes they have infected and the defender loses exactly this amount.
If the defending player ends the game and the attacking player has gained control of at least one node, then both players get zero reward.
Clearly, in this setting, the optimal first-move for the defender is to end the game immediately and set $Th=0$.
Thus, in our simplified environment, we also assume that if the defender ends the game when the attacker is not present, they incur a very large cost that demands $Th > 0$.

We analyze three cases of this game:
\begin{enumerate}
	\item All parameters are common knowledge and $s_A$ is observable to both players
	\item All parameters are common knowledge but $s_A$ is not observable to the attacker
	\item Parameters are hidden from the attacker and $s_A$ is not observable to the attacker
\end{enumerate}

In all cases of the game, the actions taken by the attacking player and $s_I$ are unknown to the defender.
This means that in the full knowledge setting, the parameters $p$ and $q$ of $T$ are known to all players, $s_A$ but not $s_I$ is observable by the defender, and the full state at each time step is observed by the attacker.
In the partial knowledge setting, $p$ and $q$ are known by all players, but only the defender can observe $s_A$ and only the attacker can observe $s_I$.
In the zero knowledge setting, only the defender knows $p$ and $q$ and can observe $s_A$, and the attacker can observe only $s_I$.

\subsection{Full knowledge}
In the full knowledge setting, the attacker and defender can both observe $s_A$, and know $p$, $q$, and $Th$.
Since the cost to the defender of ending the game when they are not under attack is so large, they must choose $Th$ such that the probability an attacker is present is close to 1.
Since $Th$ is known and $s_A$ is observable, the attacker-optimal strategy is to end the game when $SA = Th$.

\begin{proposition} \label{prop:threshold}
	The defender-optimal threshold, $Th$ is given by $\left\lfloor\left\lceil\frac{1}{p}\right\rceil (p + q - pq)\right\rfloor$.
	\begin{proof}
		At each time step from 1 to $t$, there is probability $q$ that an alert fires independent of any other action. 
		If the attacking player takes an action, there is probability $p$ that an alert fires.
		This lets us treat $p$ as the rate of arrival for alerts and view $X_m$ as a Bernoulli process.
		We wish to find the number of trials required until the first success, which follows the Geometric distribution and consequently, the expected number of trials before the first success is $\frac{1}{p}$.
		However, $t$, $X_n$, and $X_m$ cannot be directly observed. 
		
		Since only $s_A$ can be observed, we must find the expected number of alerts that occur at time $t = \frac{1}{p}$.
		As our events are discrete, we'll need to take the ceiling of $t$.
		The four possible outcomes of $\mathbf{1}_{X}$ are:
		\begin{align*}
		    (X_m = 0, X_n = 0), & (X_m = 1, X_n = 0), \\
		    (X_m = 0, X_n = 1), & (X_m = 1, X_n = 1)
		\end{align*}
		and all but the first outcome yields an alert. 
		We can therefore find the probability that an alert occurs at some time $\tau$ by taking the complement of the probability of the first event.
		Since $X_m$ and $X_n$ are independent, that probability is $(1-p)(1-q)$.
		So our expected number of alerts at time $t = \left\lceil\frac{1}{p}\right\rceil$ is:
		\begin{align*}
			((1-p)(1-q)) t & = (1-p)(1-q) \left\lceil\frac{1}{p}\right\rceil \\
			& = (pq - p - q + 1) \left\lceil\frac{1}{p}\right\rceil \\
			& = \left\lceil\frac{1}{p}\right\rceil (1 - (pq - p - q + 1))  \\
			& = \left\lceil\frac{1}{p}\right\rceil (1 - pq + p + q - 1)  \\
			& = \left\lceil\frac{1}{p}\right\rceil (p + q - pq) 
		\end{align*}
		We cannot guarantee that this number is an integer, and since the defender wants the minimal threshold, we seek the floor of the number of alerts and set:
		\begin{equation} \label{eqn:threshold}
		    Th = \left\lfloor\left\lceil\frac{1}{p}\right\rceil (p + q - pq)\right\rfloor
		\end{equation} 
	\end{proof}
\end{proposition}

The attacker-optimal strategy then, is dependent upon how ties are broken.
If the attacker always wins ties, they should act until $s_A = Th$.
If the defender always wins ties or if ties are broken randomly, the attacker should act until $Th-1 < s_A \leq Th$.
For simplicity, we assume that the attacker always wins ties throughout this work.

\subsection{Zero Observability}
In the second case, the parameters of the game are known to both players -- that is, $p, q$ are common knowledge.
Since $Th$ is chosen a priori and known by the defender, the attacking player can conclude the value of $Th$ for a rational opponent by Proposition~\ref{prop:threshold}.
However, $s_A$, cannot be observed by the attacker. 
The defending player sets $Th = \left\lfloor\left\lceil\frac{1}{p}\right\rceil (p + q - pq)\right\rfloor$ as in Equation~\ref{eqn:threshold}. 
The attacking player must therefore determine the maximum number of actions they can expect to take before $Th$ is reached.

\begin{proposition} \label{prop:pk_attack}
The optimal number of actions for the attacker to take when $s_A$ cannot be observed is $\frac{1 - p - q + pq}{p}$.

\begin{proof}
$p$ and $q$ being known, the attacker seeks to find the expected arrival time of the $Th$th alert generated by the Bernoulli process described by the joint distribution of $X_m$ and $X_n$.
Since the arrival rate for the joint process is $p + q - pq$ as demonstrated in Proposition~\ref{prop:threshold}, the attacker seeks to find $t$ such that they are able to maximize their actions. 
We want to know the number of trials, or actions, required before $Th$ alerts occur, which can be represented by the negative binomial distribution~\cite{casella2021statistical} with probability mass function:
\[\binom{k + r - 1}{r - 1}(1-p)^{k}p^{r}\]
Where $r$ is the number of successes, $k$ is the number of failures, $p$ is the probability of success, and $\binom{k + r - 1}{r - 1}$ is the binomial coefficient.
We call this random variable $X \sim NB(k; r, p)$.
The expected value of $X$ is given by $\frac{r(1-p)}{p}$, so the expected number of actions that can be taken by an attacker before the threshold is reached is given by:
\begin{align*} 
    t & = E[X] \\
    & = \frac{r(1-p)}{p} \\
    & = \frac{T(1 - (p + q - pq)}{p + q - pq}\\
    & = Th \frac{1 - p - q + pq}{p + q - pq}\\
    & \approx (\frac{1}{p} p + q - pq) \frac{1 - p - q + pq}{p + q - pq} \\
    & = \frac{1 - p - q + pq}{p} \numberthis \label{eqn:pk_attack}
\end{align*}
\end{proof}
\end{proposition}

The attacker, not knowing the value of $s_A$, rationally draws the same conclusion as the defender -- since the threshold is defined as a function of the Bernoulli processes by Equation~\ref{eqn:threshold}, they can operate until the expected number of alerts generated is equal to the expected threshold, given by Equation~\ref{eqn:pk_attack}.
In our simulation, this value is chosen by the attacker a priori.
Once the attacker has taken all their allocated actions, they end the game -- if they choose to take another action when the defender decides to end the game, they receive no reward.

\subsection{Unknown parameters}
In the setting where $p$ and $q$ are unknown, the attacker cannot make a rational assessment about $Th$. 
Since the attacker is still utility maximizing, we must establish some criteria by which the attacker decides whether or not to stop. 
In the zero-shot setting, where both the count of alerts and the environment parameters are unknown, the attacking agent has no way to reasonably choose a number of actions to take.
Attacker-defender interaction here would demand a more complex model, an example of which we discuss in the conclusion.
If the attacker is able to conduct multiple trials against a defender, they can use the number of actions $t$ they've taken along with a Bayesian updating rule to estimate $Th$ as a function of $t$. 

Over time, this attacker should optimize to approach both the partial knowledge and full knowledge attacker.
However, our goal here is to understand the impact of limited information on the agents, and so we choose an arbitrary small number of attempts for the experimental setting.
Our simple Bayesian learning agent is given 10 attempts to interact with the defender, and for each experiment, we return their maximum score and average win rate.

\section{Simulation and Discussion}
We evaluate the three models described over 20 evenly spaced values of $p$ and $q$ between 0.01 and 0.99, for a total of 400 combinations of parameters. 
In each model, play proceeds as follows:
\begin{enumerate}
    \item The defender calculates $Th$ according to Equation~\ref{eqn:threshold}
    \item Players take an action, and $s_I$ is incremented unless either player ends the game.
    \item Bernoulli trials are run for $X_m$ and $X_n$, returning 1 with probability $p$ and $q$, respectively.
    \item If either trial returns 1, increment $s_A$
    \item The defender and attacker choose their next action and repeat steps 2-5 until the game ends
\end{enumerate}
The defender's choice of action is always to pass unless $s_A = Th$, in which case, they end the game.
The attacker's choice of action varies between models.
In the full knowledge model, the attacker ends the game if $s_A = Th$, otherwise they attack another node.
In the partial knowledge model, the attacker chooses a lifespan based on Equation~\ref{eqn:pk_attack}. 
If $s_I$ is less than that lifespan, they attack another node; otherwise, they end the game.
In the zero knowledge model, the attacker is following the same rules as the partial knowledge model, but is trying to find the optimal lifespan -- at the end of each one of their attempts, they update their expected value of the joint distribution of $p$ and $q$, determine a new optimal lifespan, and try until they have exhausted their 10 total attempts. 
The average win rate and average score of those 10 attempts are then taken as one trial.
Each combination of $p$ and $q$ was run over 1 million trials, and the results of those trials -- the attacker score and wins -- averaged.

\begin{figure}[ht]
    \centering
    \includegraphics[width=0.45\textwidth]{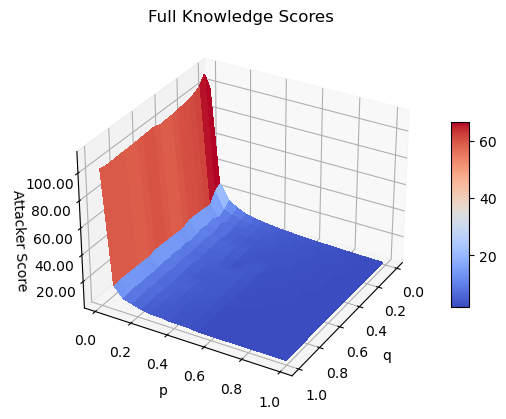}
    \caption{Average score over 1 million trials for full knowledge model}
    \label{fig:fk_score}
\end{figure}

\begin{figure} [ht]
    \centering
    \includegraphics[width=0.45\textwidth]{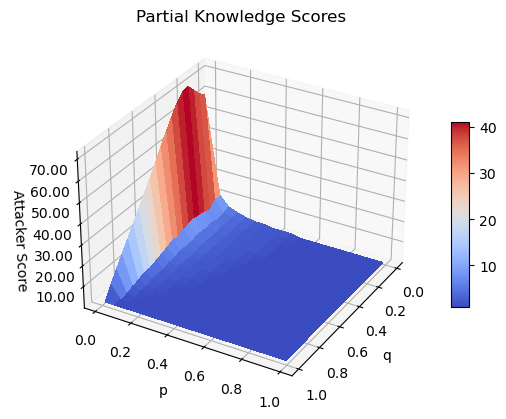}
    \caption{Average score over 1 million trials for partial knowledge model}
    \label{fig:pk_score}
\end{figure}

\begin{figure} [htbp]
    \centering
    \includegraphics[width=0.45\textwidth]{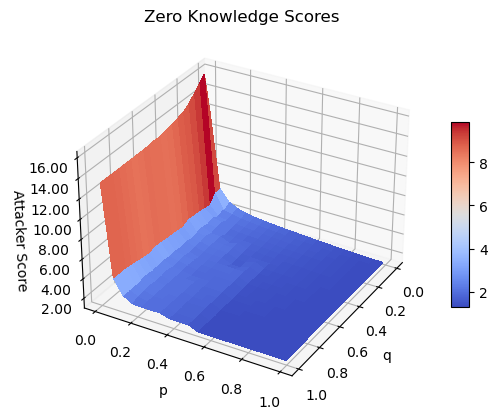}
    \caption{Average score over 1 million trials for zero knowledge model}
    \label{fig:zk_score}
\end{figure}

Figures~\ref{fig:fk_score}, \ref{fig:pk_score}, and \ref{fig:zk_score} show the average score of each model over 1 million trials as a function of $p$ and $q$.
Within these figures, we can see two emergent phenomena.
First, we can see that the general shape of the surface is the same across all three models, peaking, as we would expect, when both $p$ and $q$ are very low. 
Second, looking at specifically the index of the Z-axis, we see that the highest attained value decreases monotonically as more information is removed from the attacker. 

In the partial knowledge trials shown in Figure~\ref{fig:pk_score}, there is a curious dip in the score achieved when both $p$ and $q$ are very low.
Looking closely at both the scores and win rates shows that this is because while the win rate is very high in the partial knowledge setting in general, there is a win rate decrease at very low detection rates.
Despite boasting an average win rate of .9823 across all parameter settings and trials, the partial knowledge attacker's win rate at the lowest rate of alerting, $p = q = 0.01$ is only .41012.
This occurs due to the high number of actions that the attacker expects to achieve in the average case combined with the large variance that is observed 
when $p$ and $q$ are low. 

The first phenomenon, that the achieved score is a relatively direct function of the alert rates, follows directly from our assumptions about the threshold and how attackers gain rewards: the longer an attacker is present, the higher their potential reward; if the rate of alerts is higher, then the attacker has less time to accomplish whatever task they aim to. 
Although the values of the attacker's score where $p$ and $q$ are larger are still usually greater than zero, they are dwarfed by the values achieved when the probability of alerts is very low. 
We can also observe, by looking at the X-axis of the figures, that the value of $p$ is much more meaningful than the value of $q$.
This makes sense, since $p$ is the parameter with the largest effect in both Equations~\ref{eqn:threshold} and \ref{eqn:pk_attack}.

More interesting is the fact that for most values of $p$, the impact of $q$ is meaningful only when the attacker's knowledge is limited. 
Specifically, in the full knowledge scenario, the difference between some fixed $p$ across all values of $q$ is small.
However, as is most obvious in Figure~\ref{fig:pk_score}, the impact of a high $q$ when $p$ is low significantly changes the expectations of an attacker to persist in an environment -- information that a savvy defender could potentially exploit.
 
The second phenomenon is most pronounced by looking at the Z-axis across the three figures, which differs meaningfully across them. 
Examining the values, we find that across all parameters and trials, the full-knowledge attacker attains a mean score of 9.4701, the partial knowledge attacker attains a mean score of 3.8224, and the zero knowledge attacker attains a mean score of 2.3751. 
At the high end, the full knowledge attacker attains a maximum score of 113.0872, the partial knowledge attacker attains a maximum score of 72.0665, and the zero knowledge attacker attains a maximum score of 16.2920. 
The value of additional information to the attacker is therefore quite substantial.

\section{Conclusion and Future Work} \label{sec:conclusion}
This work considers a partially observable stochastic Bayesian game that offers useful insights even under strong assumptions about the state and action space.
In particular, we can quantify the value of total alert rate versus the true positive rate, finding that the value of an alert, from an defender's perspective, is almost entirely about how effective it is at detecting attacker activity, while overall alert rate is essentially meaningless to the attacker, except at the extremes where the overall alert rate far exceeds the true positive rate.
This is most pronounced in the scenarios where an attacker's knowledge about the parameters and the state of the world are limited -- scenarios that map more closely to real-life.
An extension of this work where the state and action spaces were expanded would offer more useful parallels and insights about these phenomena.

In this work, we have found optimal solutions for our particular limited state and action spaces. 
Since Harsanyi-Bellman ad hoc coordination depends heavily on observing the history and there is substantial uncertainty about how to choose a strategy or predict what actions an opponent may have taken -- particularly since any attacker action having been taken would demand a defender response and any defender response would end the game -- a generalization of this concept to the partially observable case of this game is difficult to assess in our limited game.
Future work in this area will generalize the HBA solution concept to arbitrary partially observable SBGs.

This work has also demonstrated that an attacker's potential success in an environment is monotonically linked to their knowledge of the environment.
Crucially, this means that when we are emulating attackers or building game theoretic models of attack and defense, the assumptions about what an attacker knows are nontrivial, and making incorrect assumptions about attacker knowledge and behavior can have negative impacts on defender optimization -- to wit, if we assume that attackers are omnipotent, then our optimal defensive strategy is very aggressive, while an optimal defensive strategy under better assumptions may yield more uptime for users.
Here, our defender admits only a fixed strategy against an attacker for whom they have no options.
However, a more dynamic defender would be able to anticipate an attacker's actions given their knowledge, and optimize against that. 

Finally, our zero-knowledge attacker is given a small, arbitrary number of opportunities to learn the environment and optimize their score. 
In order to conduct a more reasonable few-shot approach to the zero-knowledge environment, we should consider a model with a larger state and action space.
In order for the attacker to be a learning agent, some local feedback -- feedback aside from winning or losing the game -- would allow for a more robust interplay and allow for the incorporation of attacker learning dynamics. 
Given the extant literature on the use of reinforcement learning for analogous problems, we would seek to apply that same technique in few-shot settings of a stochastic Bayesian game.


\bibliography{references}

\end{document}